\documentclass[12pt,reqno]{amsart}
\usepackage{graphicx}
\usepackage{amscd}
\usepackage{amsmath}
\usepackage{epsfig}
\usepackage{amsfonts}
\usepackage{amssymb}

\providecommand{\U}[1]{\protect\rule{.1in}{.1in}}
\providecommand{\U}[1]{\protect\rule{.1in}{.1in}}
\allowdisplaybreaks

\theoremstyle{plain}

\newtheorem{lemma}{Lemma}

\newtheorem{remark}{Remark}

\numberwithin{equation}{section}

\input{tcilatex}

\begin{document}
\title[Dynamical Invariants and Berry Phase]{Dynamical Invariants and
Berry's Phase for Generalized Driven Harmonic Oscillators}
\author{Barbara Sanborn }
\address{Department of Mathematics, Western Washington University,
Bellingham, WA 98225-9063, U.S.A.}
\email{barbara.sanborn@wwu.edu}
\author{Sergei K. Suslov}
\address{School of Mathematical and Statistical Sciences \& Mathematical,
Computational and Modeling Sciences Center, Arizona State University, Tempe,
AZ 85287--1804, U.S.A.}
\email{sks@asu.edu}
\urladdr{http://hahn.la.asu.edu/\symbol{126}suslov/index.html}
\author{Luc Vinet}
\address{Centre de Recherches Math\'{e}matiques, Universit\'{e} de Montr\'{e}%
al, Montr\'{e}al, Qu\'{e}bec, Centre-ville Station, P.O. Box 6128, Canada
H3C 3J7.}
\email{luc.vinet@umontreal.ca}
\date{\today }

\begin{abstract}
We present quadratic dynamical invariant and evaluate Berry's phase for the
time-dependent Schr\"{o}dinger equation with the most general variable
quadratic Hamiltonian.
\end{abstract}

\maketitle

\section{Introduction}

In the previous Letter \cite{Lan:Lop:Sus}, the exact wave functions for
generalized (driven) harmonic oscillators \cite{Berry85}, \cite%
{Cor-Sot:Lop:Sua:Sus}, \cite{Hannay85}, \cite{Leach90}, \cite{Lo93}, \cite%
{Wolf81}, \cite{XuGao96}, \cite{Yeon:Lee:Um:George:Pandey93} have been
constructed in terms of Hermite polynomials by transforming the
time-dependent Schr\"{o}dinger equation into an autonomous form \cite%
{Zhukov99}. Relationships with certain Ermakov and Riccati-type systems have
been investigated. A goal of this Letter is to find the corresponding
dynamical invariants and to evaluate Berry's phase \cite{Berry84}, \cite%
{Berry85}, \cite{Simon83}, \cite{WilczekZee84} for quantum systems with
general variable quadratic Hamiltonians as an extension of the works \cite%
{CerLeja89}, \cite{EngGh88}, \cite{GaoXuQian90}, \cite{Kobe90}, \cite{Kobe91}%
, \cite{Leach90}, \cite{Moral88}, \cite{MontKorNu94}, \cite{Suslov10} (see
also references therein).

\section{Generalized Driven Harmonic Oscillators}

We consider the one-dimensional time-dependent Schr\"{o}dinger equation%
\begin{equation}
i\frac{\partial \psi }{\partial t}=H\psi ,  \label{Schroedinger}
\end{equation}%
where the variable Hamiltonian $H=Q\left( p,x\right) $ is an arbitrary
quadratic of two operators $p=-i\partial /\partial x$ and $x,$ namely,%
\begin{equation}
i\psi _{t}=-a\left( t\right) \psi _{xx}+b\left( t\right) x^{2}\psi -ic\left(
t\right) x\psi _{x}-id\left( t\right) \psi -f\left( t\right) x\psi +ig\left(
t\right) \psi _{x}  \label{SchroedingerQuadratic}
\end{equation}%
($a,$ $b,$ $c,$ $d,$ $f$ and $g$ are suitable real-valued functions of time
only). We shall refer to these quantum systems as the \textit{generalized
(driven) harmonic oscillators}. A general approach and known elementary
solutions can be found in Refs.~\cite{Cor-Sot:Lop:Sua:Sus}, \cite%
{Cor-Sot:Sua:Sus}, \cite{Cor-Sot:Sua:SusInv}, \cite{Cor-Sot:Sus}, \cite%
{Dod:Mal:Man75}, \cite{FeynmanPhD}, \cite{Feynman}, \cite{Fey:Hib}, \cite%
{Lan:Lop:Sus}, \cite{Lo93}, \cite{Lop:Sus}, \cite{Me:Co:Su}, \cite{Suaz:Sus}%
, \cite{Wolf81} and \cite{Yeon:Lee:Um:George:Pandey93}. In addition, a case
related to Airy functions is discussed in \cite{Lan:Sus} and Ref.~\cite%
{Cor-Sot:SusDPO} deals with another special case of transcendental solutions.

In this Letter, we shall use the following result established in \cite%
{Lan:Lop:Sus}.

\begin{lemma}
The substitution%
\begin{equation}
\psi =\frac{e^{i\left( \alpha \left( t\right) x^{2}+\delta \left( t\right)
x+\kappa \left( t\right) \right) }}{\sqrt{\mu \left( t\right) }}\ \chi
\left( \xi ,\tau \right) ,\qquad \xi =\beta \left( t\right) x+\varepsilon
\left( t\right) ,\quad \tau =\gamma \left( t\right)  \label{Ansatz}
\end{equation}%
transforms the non-autonomous and inhomogeneous Schr\"{o}dinger equation (%
\ref{SchroedingerQuadratic}) into the autonomous form%
\begin{equation}
-i\chi _{\tau }=-\chi _{\xi \xi }+c_{0}\xi ^{2}\chi \qquad \left(
c_{0}=0,1\right)  \label{ASEq}
\end{equation}%
provided that%
\begin{equation}
\frac{d\alpha }{dt}+b+2c\alpha +4a\alpha ^{2}=c_{0}a\beta ^{4},  \label{SysA}
\end{equation}%
\begin{equation}
\frac{d\beta }{dt}+\left( c+4a\alpha \right) \beta =0,  \label{SysB}
\end{equation}%
\begin{equation}
\frac{d\gamma }{dt}+a\beta ^{2}=0  \label{SysC}
\end{equation}%
and%
\begin{equation}
\frac{d\delta }{dt}+\left( c+4a\alpha \right) \delta =f+2g\alpha
+2c_{0}a\beta ^{3}\varepsilon ,  \label{SysD}
\end{equation}%
\begin{equation}
\frac{d\varepsilon }{dt}=\left( g-2a\delta \right) \beta ,  \label{SysE}
\end{equation}%
\begin{equation}
\frac{d\kappa }{dt}=g\delta -a\delta ^{2}+c_{0}a\beta ^{2}\varepsilon ^{2}.
\label{SysF}
\end{equation}%
Here%
\begin{equation}
\alpha =\frac{1}{4a}\frac{\mu ^{\prime }}{\mu }-\frac{d}{2a}.  \label{Alpha}
\end{equation}
\end{lemma}

The substitution (\ref{Alpha}) reduces the inhomogeneous equation (\ref{SysA}%
) to the second order ordinary differential equation%
\begin{equation}
\mu ^{\prime \prime }-\tau \left( t\right) \mu ^{\prime }+4\sigma \left(
t\right) \mu =c_{0}\left( 2a\right) ^{2}\beta ^{4}\mu ,  \label{CharEq}
\end{equation}%
that has the familiar time-varying coefficients%
\begin{equation}
\tau \left( t\right) =\frac{a^{\prime }}{a}-2c+4d,\qquad \sigma \left(
t\right) =ab-cd+d^{2}+\frac{d}{2}\left( \frac{a^{\prime }}{a}-\frac{%
d^{\prime }}{d}\right) .  \label{TauSigma}
\end{equation}

When $c_{0}=0$, equation (\ref{SysA}) is called the \textit{Riccati
nonlinear differential equation} \cite{Wa}, \cite{Wh:Wa} and the system (\ref%
{SysA})--(\ref{SysF}) shall be referred to as a \textit{Riccati-type system}%
. (Similar terminology is used in \cite{SuazoSusVega10}, \cite%
{SuazoSusVega11} for the corresponding parabolic equation.) If $c_{0}=1,$\
equation (\ref{CharEq}) can be reduced to a generalized version of the 
\textit{Ermakov nonlinear differential equation} (see, for example, \cite%
{Cor-Sot:Sua:SusInv}, \cite{Ermakov}, \cite{Leach:Andrio08}, \cite{Suslov10}
and references therein regarding Ermakov's equation) and we shall refer to
the corresponding system (\ref{SysA})--(\ref{SysF}) with $c_{0}\neq 0$ as an 
\textit{Ermakov-type system}. Throughout this Letter, we use the notations
from Ref.~\cite{Lan:Lop:Sus} where a more detailed bibliography on the
quadratic systems can be found.

Using standard oscillator wave functions for equation (\ref{ASEq}) when $%
c_{0}=1$ (for example, \cite{Flu}, \cite{La:Lif} and/or \cite{Merz}) results
in the solution%
\begin{equation}
\psi _{n}\left( x,t\right) =\frac{e^{i\left( \alpha x^{2}+\delta x+\kappa
\right) +i\left( 2n+1\right) \gamma }}{\sqrt{2^{n}n!\mu \sqrt{\pi }}}\
e^{-\left( \beta x+\varepsilon \right) ^{2}/2}\ H_{n}\left( \beta
x+\varepsilon \right) ,  \label{WaveFunction}
\end{equation}%
where $H_{n}\left( x\right) $ are the Hermite polynomials \cite{Ni:Su:Uv}
and the general real-valued solution of the Ermakov-type system (\ref{SysA}%
)--(\ref{SysF}) is available in Ref.~\cite{Lan:Lop:Sus} --- Lemma~3,
Eqs.~(42)--(48).

The Green function of generalized harmonic oscillators has been constructed
in Ref.~\cite{Cor-Sot:Lop:Sua:Sus}. (See also important previous works \cite%
{Dodonov:Man'koFIAN87}, \cite{Malkin:Man'ko79}, \cite{Wolf81}, \cite%
{Yeon:Lee:Um:George:Pandey93}, \cite{Zhukov99} and references therein for
more details.)

The corresponding Cauchy initial value problem can be solved (formally) by
the superposition principle:%
\begin{equation}
\psi \left( x,t\right) =\int_{-\infty }^{\infty }G\left( x,y,t\right) \psi
\left( y,0\right) \ dy  \label{Superposition}
\end{equation}%
for some suitable initial data $\psi \left( x,0\right) =\varphi \left(
x\right) $ (see Refs.~\cite{Cor-Sot:Lop:Sua:Sus}, \cite{Suaz:Sus} and \cite%
{Suslov10} for further details). The corresponding eigenfunction expansion
can be written in terms of the wave functions (\ref{WaveFunction}) as follows%
\begin{equation}
\psi \left( x,t\right) =\sum_{n=0}^{\infty }c_{n}\ \psi _{n}\left(
x,t\right) ,  \label{EigenfunctionExp}
\end{equation}%
where the time-independent coefficients are given by%
\begin{equation}
c_{n}=\dfrac{\dint_{-\infty }^{\infty }\psi _{n}^{\ast }\left( x,t\right)
\psi \left( x,0\right) \ dx}{\dint_{-\infty }^{\infty }\left\vert \psi
_{n}\left( x,0\right) \right\vert ^{2}\ dx}.  \label{EigenfunctionCoeff}
\end{equation}%
This expansion complements the integral form of solution (\ref{Superposition}%
).

The maximum symmetry group of the autonomous Schr\"{o}dinger equation (\ref%
{ASEq}) is studied in \cite{Niederer72} and \cite{Niederer73} (see also \cite%
{VinetZhedanov2011} and references therein).

\section{Dynamical Invariants for Generalized Driven Harmonic Oscillators}

A concept of dynamical invariants for generalized harmonic oscillators has
been recently revisited in Refs.~\cite{Cor-Sot:Sua:SusInv} and \cite%
{Suslov10} (see \cite{Dodonov2000}, \cite{Dod:Mal:Man75}, \cite%
{Dodonov:Man'koFIAN87}, \cite{Malkin:Man'ko79}, \cite{Malk:Man:Trif73} and
references therein for classical works). In this Letter, we would like to
point out a simple extension of the quadratic dynamical invariant to the
case of driven oscillators:%
\begin{eqnarray}
E\left( t\right) &=&\frac{\lambda \left( t\right) }{2}\left[ \widehat{a}%
\left( t\right) \widehat{a}^{\dagger }\left( t\right) +\widehat{a}^{\dagger
}\left( t\right) \widehat{a}\left( t\right) \right]  \label{QInvariant} \\
&=&\frac{\lambda \left( t\right) }{2}\left[ \frac{\left( p-2\alpha x-\delta
\right) ^{2}}{\beta ^{2}}+\left( \beta x+\varepsilon \right) ^{2}\right]
,\qquad \frac{d}{dt}\langle E\rangle =0.  \notag
\end{eqnarray}%
(See also \cite{EngGh88}, \cite{GaoXuQian90}, \cite{MontKorNu94} and \cite%
{XuGao96}.) Here, $\lambda \left( t\right) =\exp \left( -\dint_{0}^{t}\left(
c\left( s\right) -2d\left( s\right) \right) \ ds\right) $ and the
corresponding time-dependent annihilation $\widehat{a}\left( t\right) $ and
creation $\widehat{a}^{\dagger }\left( t\right) $ operators are explicitly
given by%
\begin{eqnarray}
\widehat{a}\left( t\right) &=&\frac{1}{\sqrt{2}}\left( \beta x+\varepsilon +i%
\frac{p-2\alpha x-\delta }{\beta }\right) ,  \label{a(t)} \\
\widehat{a}^{\dagger }\left( t\right) &=&\frac{1}{\sqrt{2}}\left( \beta
x+\varepsilon -i\frac{p-2\alpha x-\delta }{\beta }\right)  \label{across(t)}
\end{eqnarray}%
with $p=i^{-1}\partial /\partial x$ in terms of solutions of the
Ermarov-type system (\ref{SysA})--(\ref{SysF}). These operators satisfy the
canonical commutation relation:%
\begin{equation}
\widehat{a}\left( t\right) \widehat{a}^{\dagger }\left( t\right) -\widehat{a}%
^{\dagger }\left( t\right) \widehat{a}\left( t\right) =1.
\label{commutatora(t)across(t)}
\end{equation}%
The oscillator-type spectrum of the dynamical invariant $E$ can be obtained
in a standard way by using the Heisenberg--Weyl algebra of the rasing and
lowering operators (a \textquotedblleft second
quantization\textquotedblright\ \cite{Lewis:Riesen69}, the Fock states):%
\begin{equation}
\widehat{a}\left( t\right) \Psi _{n}\left( x,t\right) =\sqrt{n}\ \Psi
_{n-1}\left( x,t\right) ,\quad \widehat{a}^{\dagger }\left( t\right) \Psi
_{n}\left( x,t\right) =\sqrt{n+1}\ \Psi _{n+1}\left( x,t\right) ,
\label{annandcratoperactions}
\end{equation}%
\begin{equation}
E\left( t\right) \Psi _{n}\left( x,t\right) =\lambda \left( t\right) \left(
n+\frac{1}{2}\right) \Psi _{n}\left( x,t\right) .  \label{Eeigenvp}
\end{equation}%
The corresponding orthogonal time-dependent eigenfunctions are given by%
\begin{equation}
\Psi _{n}\left( x,t\right) =\frac{e^{i\left( \alpha x^{2}+\delta x+\kappa
\right) -\left( \beta x+\varepsilon \right) ^{2}/2}}{\sqrt{2^{n}n!\mu \sqrt{%
\pi }}}\ H_{n}\left( \beta x+\varepsilon \right) ,\qquad \left\langle \Psi
_{m},\Psi _{n}\right\rangle =\delta _{mn}\ \lambda ^{-1}  \label{Eeigenfs}
\end{equation}%
(provided that $\beta \left( 0\right) \mu \left( 0\right) =1,$ when $\beta
\mu =\lambda $ \cite{Lan:Lop:Sus}) in terms of Hermite polynomials \cite%
{Ni:Su:Uv} and%
\begin{equation}
\psi _{n}\left( x,t\right) =e^{i\left( 2n+1\right) \gamma \left( t\right) }\
\Psi _{n}\left( x,t\right)  \label{WaveInvariantFunctions}
\end{equation}%
is the relation to the wave functions (\ref{WaveFunction}) with%
\begin{equation}
\varphi _{n}\left( t\right) =-\left( 2n+1\right) \gamma \left( t\right)
\label{LewisPhase}
\end{equation}%
being the Lewis phase \cite{GaoXuQian90}, \cite{Leach90}, \cite%
{Lewis:Riesen69}.

The dynamic invariant operator derivative identity \cite{Cor-Sot:Sua:SusInv}%
, \cite{Suslov10}:%
\begin{equation}
\frac{dE}{dt}=\frac{\partial E}{\partial t}+i^{-1}\left( EH-H^{\dagger
}E\right) =0  \label{InvariantDer}
\end{equation}%
can be veryfied in the following fashion. Introducing new linear momentum
and coordinate operators in the form%
\begin{equation}
P=\frac{\lambda }{\beta }\left( p-2\alpha x-\delta \right) ,\qquad Q=\lambda
\left( \beta x+\varepsilon \right) ,  \label{PQoperators}
\end{equation}%
when $\left[ Q,P\right] =i\lambda ^{2}$ (a generalized canonical
transformation), one can derive the simple differentiation rules%
\begin{equation}
\frac{dP}{dt}=-2c_{0}a\beta ^{2}Q,\qquad \frac{dQ}{dt}=2a\beta ^{2}P.
\label{PQDiff}
\end{equation}%
(It is worth noting that if $c_{0}=0,$ the operator $P$ becomes the linear
invariant of Dodonov, Malkin, Manko and Trifonov \cite{Dod:Mal:Man75}, \cite%
{Dodonov:Man'koFIAN87}, \cite{Malk:Man:Trif73}, \cite{XuGao96} for
generalized driven harmonic oscillators.)

Then%
\begin{equation}
E=\frac{\lambda ^{-1}}{2}\left( P^{2}+c_{0}Q^{2}\right) \qquad \left(
c_{0}=0,1\right)  \label{EPQinvariant}
\end{equation}%
and it is useful to realise that $E$ is just the original Hamiltonian $H$
after the canonical transformation \cite{Leach90}. The required operator
indentity (\ref{InvariantDer}) can be formally derived with the aid of
product rule (3.7) of Ref.~\cite{Suslov10} (quantum calculus):%
\begin{eqnarray}
2\frac{dE}{dt} &=&\frac{d}{dt}\left( \lambda ^{-1}P^{2}\right) +c_{0}\frac{d%
}{dt}\left( \lambda ^{-1}Q^{2}\right) \\
&=&\lambda ^{-1}\left( \frac{dP}{dt}P+P\frac{dP}{dt}\right) +c_{0}\lambda
^{-1}\left( \frac{dQ}{dt}Q+Q\frac{dQ}{dt}\right)  \notag
\end{eqnarray}%
and by (\ref{PQDiff}):%
\begin{equation}
\lambda \frac{dE}{dt}=c_{0}a\beta ^{2}\left( -QP-PQ+PQ+QP\right) =0,
\end{equation}%
which completes the proof.

\begin{remark}
The kernel%
\begin{equation}
K\left( x,y,t\right) =\frac{1}{\sqrt{\mu }}e^{i\left( \alpha x^{2}+\beta
xy+\gamma y^{2}+\delta x+\varepsilon y+\kappa \right) }  \label{Kkernel}
\end{equation}%
is a particular solution of the Schr\"{o}dinger equation (\ref%
{SchroedingerQuadratic}) for any solution of the Riccati-type system (\ref%
{SysA})--(\ref{Alpha}) with $c_{0}=0$ \cite{Cor-Sot:Lop:Sua:Sus}. A direct
calculation shows that this kernel is an eigenfunction%
\begin{equation}
\beta ^{-1}\left( p-2\alpha x-\delta \right) K\left( x,y,t\right) =yK\left(
x,y,t\right)  \label{KEigen}
\end{equation}%
of the linear dynamical invariant \cite{Suslov10}.
\end{remark}

\section{Evaluation of Berry's Phase}

The holonomic effect in quantum mechanics known as Berry's phase \cite%
{Berry84}, \cite{Berry85} had received considerable attention over the years
(see, for example, \cite{CerLeja89}, \cite{EngGh88}, \cite{Gerry89}, \cite%
{Hannay85}, \cite{Kobe90}, \cite{Kobe91}, \cite{Leach90}, \cite{MontKorNu94}%
, \cite{Moral88}, \cite{MontKorNu94}, \cite{Simon83}, \cite{Vinet88}, \cite%
{WilczekZee84}, \cite{Xiaoetal10} and references therein). The solution of
the time-dependent Schr\"{o}dinger equation (\ref{SchroedingerQuadratic})
has the form (\ref{EigenfunctionExp}) with $\ $the oscillator-type wave
functions $\psi _{n}\left( x,t\right) $ presented by (\ref{WaveFunction}) 
\cite{Lan:Lop:Sus}:%
\begin{equation}
\psi _{n}\left( x,t\right) =e^{-i\varphi _{n}\left( t\right) }\ \Psi
_{n}\left( x,t\right) ,  \label{psiPsi}
\end{equation}%
where $\varphi _{n}\left( t\right) $ is the Lewis (or dynamical) phase and $%
\Psi _{n}\left( x,t\right) $ is the eigenfunction of quadratic invariant (%
\ref{Eeigenvp}). (In the self-adjoint case, one chooses $c=2d$ when $\lambda
=1.$)

Then%
\begin{equation}
i\int_{%
\mathbb{R}
}\psi _{n}^{\ast }\frac{\partial \psi _{n}}{\partial t}\ dx=\lambda ^{-1}%
\frac{d\varphi _{n}}{dt}+i\int_{%
\mathbb{R}
}\Psi _{n}^{\ast }\frac{\partial \Psi _{n}}{\partial t}\ dx
\label{LewisBerryPhases}
\end{equation}%
and Berry's phase $\theta _{n}$ is given by%
\begin{equation}
\lambda ^{-1}\frac{d\theta _{n}}{dt}=\func{Re}\left( i\int_{%
\mathbb{R}
}\Psi _{n}^{\ast }\frac{\partial \Psi _{n}}{\partial t}\ dx\right) =\func{Re}%
\left( i\left\langle \Psi _{n},\frac{\partial }{\partial t}\Psi
_{n}\right\rangle \right) .  \label{BerryPhase}
\end{equation}%
Here, the eigenfunction $\Psi _{n}$ is a $\gamma $-free part \cite{Leach90}
of the wave function (\ref{WaveFunction}), namely%
\begin{equation}
\Psi _{n}=\lambda ^{-1/2}e^{i\left( \alpha x^{2}+\delta x+\kappa \right)
}\Phi _{n}\left( x,t\right) ,  \label{WFHarmOscillator}
\end{equation}%
and $\Phi _{n}$ is, essentially, the real-valued stationary orthonormal wave
function for the simple harmonic oscillator with respect to the new variable 
$\xi =\beta x+\varepsilon $ (see (\ref{Eeigenfs}) and (\ref{HONorm})). The
integral (\ref{BerryPhase}) can be evaluated as in Refs.~\cite{Kobe91} and 
\cite{Leach90}:%
\begin{eqnarray*}
&&\lambda \left\langle \Psi _{n},\frac{\partial \Psi _{n}}{\partial t}%
\right\rangle =i\left\langle \Phi _{n},\left( \frac{d\alpha }{dt}x^{2}+\frac{%
d\delta }{dt}x+\frac{d\kappa }{dt}\right) \Phi _{n}\right\rangle +\frac{1}{2}%
\left( c-2d\right) +\left\langle \Phi _{n},\frac{\partial \Phi _{n}}{%
\partial t}\right\rangle \\
&&\qquad =i\frac{d\alpha }{dt}\left\langle \Phi _{n},x^{2}\Phi
_{n}\right\rangle +i\frac{d\delta }{dt}\left\langle \Phi _{n},x\Phi
_{n}\right\rangle +i\frac{d\kappa }{dt}\left\langle \Phi _{n},\Phi
_{n}\right\rangle +\frac{1}{2}\left( c-2d\right) +\left\langle \Phi _{n},%
\frac{\partial \Phi _{n}}{\partial t}\right\rangle ,
\end{eqnarray*}%
where the last term is zero due to the normalization condition%
\begin{equation}
\int_{-\infty }^{\infty }\Phi _{n}^{2}\ dx=1.  \label{HONorm}
\end{equation}%
Moreover,%
\begin{eqnarray*}
\left\langle \Phi _{n},x^{2}\Phi _{n}\right\rangle &=&\beta
^{-3}\int_{-\infty }^{\infty }\left( \xi ^{2}+\varepsilon ^{2}\right) \Phi
_{n}^{2}\ d\xi =\beta ^{-2}\left( \varepsilon ^{2}+n+\frac{1}{2}\right) , \\
\left\langle \Phi _{n},x\Phi _{n}\right\rangle &=&-\varepsilon \beta
^{-2}\int_{-\infty }^{\infty }\Phi _{n}^{2}\ d\xi =-\varepsilon \beta ^{-1}
\end{eqnarray*}%
with the help of%
\begin{equation}
\beta ^{-1}\int_{-\infty }^{\infty }\xi \Phi _{n}^{2}\ d\xi =0,\qquad \beta
^{-1}\int_{-\infty }^{\infty }\xi ^{2}\Phi _{n}^{2}\ d\xi =n+\frac{1}{2}.
\label{MatElems}
\end{equation}%
As a result,%
\begin{equation}
\frac{d\theta _{n}}{dt}=-\beta ^{-2}\left( \varepsilon ^{2}+n+\frac{1}{2}%
\right) \frac{d\alpha }{dt}+\varepsilon \beta ^{-1}\frac{d\delta }{dt}-\frac{%
d\kappa }{dt}  \label{BerryGamma}
\end{equation}%
and the phase $\theta _{n}$ can be obtain by integrating (\ref{BerryGamma}).
Our observation reveals the connection of Berry's phase with the
Ermakov-type system (\ref{SysA})--(\ref{Alpha}), whose general solution is
found in Ref.~\cite{Lan:Lop:Sus}.

When $c-2d=f=g=0,$ one may choose $\delta =\varepsilon =\kappa =0$ and our
expression (\ref{BerryGamma}) simplifies to%
\begin{eqnarray}
\dfrac{d\theta _{n}}{dt} &=&-\mu ^{2}\left( n+\frac{1}{2}\right) \frac{%
d\alpha }{dt}  \label{GammaSimple} \\
&=&-\frac{1}{4a}\left( n+\frac{1}{2}\right) \left[ \mu ^{\prime \prime }\mu
-\left( \mu ^{\prime }\right) ^{2}-\frac{a^{\prime }}{a}\mu ^{\prime }\mu
+2d\left( \frac{a^{\prime }}{a}-\frac{d^{\prime }}{d}\right) \mu ^{2}\right]
\notag
\end{eqnarray}%
with the help of (\ref{Alpha}). The function $\mu $ is a solution of the
Ermakov equation (\ref{CharEq})--(\ref{TauSigma}) with $c_{0}=1$ and $\beta
=\mu ^{-1}.$ This result is consistent with Refs.~\cite{EngGh88} and \cite%
{Leach90}, where the original expression of Ref.~\cite{Moral88} has been
corrected.

\section{An Alternative Derivation of Berry's Phase}

In view of (\ref{Schroedinger}) and (\ref{psiPsi})--(\ref{BerryPhase}), we
get%
\begin{equation}
\lambda ^{-1}\left( \frac{d\theta _{n}}{dt}+\frac{d\varphi _{n}}{dt}\right) =%
\func{Re}\left\langle \psi _{n},H\psi _{n}\right\rangle =\func{Re}%
\left\langle \Psi _{n},H\Psi _{n}\right\rangle ,  \label{ThetaPhiH}
\end{equation}%
because the Hamiltonian in (\ref{Schroedinger})--(\ref{SchroedingerQuadratic}
does not involve time differentiation. Here,%
\begin{equation}
H=ap^{2}+bx^{2}+\frac{c}{2}\left( px+xp\right) +\frac{i}{2}\left(
c-2d\right) -fx-gp  \label{Hpx}
\end{equation}%
and the position and linear momentum operators are given by%
\begin{eqnarray}
x &=&\frac{1}{\beta }\left[ \frac{1}{\sqrt{2}}\left( \widehat{a}+\widehat{a}%
^{\dagger }\right) -\varepsilon \right] ,  \label{x2a} \\
p &=&\frac{\beta }{i\sqrt{2}}\left( \widehat{a}-\widehat{a}^{\dagger
}\right) +\frac{\sqrt{2}\alpha }{\beta }\left( \widehat{a}+\widehat{a}%
^{\dagger }\right) +\delta -\frac{2\alpha \varepsilon }{\beta }  \label{p2a}
\end{eqnarray}%
in terms of the creation and annihilation operators (\ref{a(t)})--(\ref%
{across(t)}). After the substitution, the Hamiltonian takes the form%
\begin{eqnarray}
H &=&\left[ \frac{a}{2}\left( \frac{4\alpha ^{2}}{\beta ^{2}}-\beta
^{2}\right) +\frac{b+2c\alpha }{\beta ^{2}}-\frac{i}{2}\left( c+4a\alpha
\right) \right] \left( \widehat{a}\right) ^{2}  \label{H2a1a} \\
&&+\left[ \frac{a}{2}\left( \frac{4\alpha ^{2}}{\beta ^{2}}-\beta
^{2}\right) +\frac{b+2c\alpha }{\beta ^{2}}+\frac{i}{2}\left( c+4a\alpha
\right) \right] \left( \widehat{a}^{\dagger }\right) ^{2}  \notag \\
&&+\frac{1}{2}\left[ a\left( \beta ^{2}+\frac{4\alpha ^{2}}{\beta ^{2}}%
\right) +\frac{b+2c\alpha }{\beta ^{2}}\right] \left( \widehat{a}\widehat{a}%
^{\dagger }+\widehat{a}^{\dagger }\widehat{a}\right) +\frac{i}{2}\left(
c-2d\right)  \notag \\
&&+\sqrt{2}\left[ \frac{4a\alpha +c}{2\beta }\left( \delta -\frac{2\alpha
\varepsilon }{\beta }\right) -\frac{\varepsilon }{\beta ^{2}}\left(
b+c\alpha \right) -\frac{f+2g\alpha }{2\beta }\right.  \notag \\
&&\quad \quad +\left. i\left( \frac{\beta }{2}\left( g-2a\delta \right) +%
\frac{\varepsilon }{2}\left( c+4a\alpha \right) \right) \right] \widehat{a} 
\notag \\
&&+\sqrt{2}\left[ \frac{4a\alpha +c}{2\beta }\left( \delta -\frac{2\alpha
\varepsilon }{\beta }\right) -\frac{\varepsilon }{\beta ^{2}}\left(
b+c\alpha \right) -\frac{f+2g\alpha }{2\beta }\right.  \notag \\
&&\quad \quad -\left. i\left( \frac{\beta }{2}\left( g-2a\delta \right) +%
\frac{\varepsilon }{2}\left( c+4a\alpha \right) \right) \right] \widehat{a}%
^{\dagger }  \notag \\
&&+a\left( \delta -\frac{2\alpha \varepsilon }{\beta }\right) ^{2}+\frac{%
\varepsilon }{\beta }\left( f+\frac{b\varepsilon }{\beta }\right) -\left(
\delta -\frac{2\alpha \varepsilon }{\beta }\right) \left( g+\frac{%
c\varepsilon }{\beta }\right) .  \notag
\end{eqnarray}%
Here,%
\begin{equation}
J_{+}=\frac{1}{2}\left( \widehat{a}^{\dagger }\right) ^{2},\qquad J_{-}=%
\frac{1}{2}\left( \widehat{a}\right) ^{2},\qquad J_{0}=\frac{1}{4}\left( 
\widehat{a}\widehat{a}^{\dagger }+\widehat{a}^{\dagger }\widehat{a}\right)
\label{SU(1,1)generators}
\end{equation}%
are the generators of a non-compact $SU\left( 1,1\right) $ algebra:%
\begin{equation}
\left[ J_{0},J_{\pm }\right] =\pm J_{\pm },\qquad \left[ J_{+},J_{-}\right]
=-2J_{0}  \label{SU(1,1)commutators}
\end{equation}%
and, therefore, a use can be made of the group properties of the
corresponding discrete positive series $\mathcal{D}_{+}^{j}$ for further
investigation of Berry's phase. (This is a `standard procedure' for
quadratic Hamiltonians --- more details can be found in Refs.~\cite%
{CerLeja89}, \cite{Gerry89}, \cite{Lo93}, \cite{Malkin:Man'ko79}, \cite%
{Me:Co:Su}, \cite{Ni:Su:Uv}, \cite{Smir:Shit}, \cite{Vinet88} and/or
elsewhere.) Together, the linears and bilinears in $\widehat{a}$ and $%
\widehat{a}^{\dagger }$ realize the semi-direct sum of the $SU\left(
1,1\right) $ and the Heisenberg algebra (\ref{commutatora(t)across(t)}) (see
Ref.~\cite{VinetZhedanov2011} for more details).

Thus%
\begin{eqnarray}
\lambda \func{Re}\left\langle \Psi _{n},H\Psi _{n}\right\rangle &=&\left( n+%
\frac{1}{2}\right) \left[ a\left( \beta ^{2}+\frac{4\alpha ^{2}}{\beta ^{2}}%
\right) +\frac{b+2c\alpha }{\beta ^{2}}\right]  \label{Psi2A} \\
&&+a\left( \delta -\frac{2\alpha \varepsilon }{\beta }\right) ^{2}+\frac{%
\varepsilon }{\beta }\left( f+\frac{b\varepsilon }{\beta }\right) -\left(
\delta -\frac{2\alpha \varepsilon }{\beta }\right) \left( g+\frac{%
c\varepsilon }{\beta }\right)  \notag
\end{eqnarray}%
by (\ref{annandcratoperactions})--(\ref{Eeigenvp}).

Finally, from (\ref{LewisPhase}) and (\ref{ThetaPhiH}) we arrive at a
different formula for Berry's phase 
\begin{eqnarray}
\frac{d\theta _{n}}{dt} &=&\left( n+\frac{1}{2}\right) \left[ a\left( \frac{%
4\alpha ^{2}}{\beta ^{2}}-\beta ^{2}\right) +\frac{b+2c\alpha }{\beta ^{2}}%
\right]  \label{BerryTheta} \\
&&+a\left( \delta -\frac{2\alpha \varepsilon }{\beta }\right) ^{2}+\frac{%
\varepsilon }{\beta }\left( f+\frac{b\varepsilon }{\beta }\right) -\left(
\delta -\frac{2\alpha \varepsilon }{\beta }\right) \left( g+\frac{%
c\varepsilon }{\beta }\right) ,  \notag
\end{eqnarray}%
which is consistent with the previuous expression (\ref{BerryGamma}) for any
solution of the Ermakov-type system (\ref{SysA})--(\ref{SysF}) $\left(
c_{0}=1\right) .$

\noindent \textbf{Acknowledgments.\/} We thank Carlos Castillo-Ch\'{a}vez
and Vladimir~I.~Man'ko for valuable discussions and encouragement. One of us
(SKS) is grateful to the organizers of the $12^{\text{th}}$ ICSSUR (Foz do
Igua\c{c}u, Brazil, May 02--06, 2011) for their hospitality.

\end{document}